\documentclass[pra,aps,twocolumn,showpacs]{revtex4}

\usepackage{amsmath}
\usepackage{bm}
\usepackage{graphicx}

\begin{document}

\title{
Stabilization of a Bose-Einstein droplet by hyperfine Rabi oscillations
}

\author{Hiroki Saito$^1$}
\author{Randall G. Hulet$^2$}
\author{Masahito Ueda$^{3,4}$}
\affiliation{$^1$Department of Applied Physics and Chemistry, The
University of Electro-Communications, Tokyo 182-8585, Japan \\
$^2$Department of Physics and Astronomy and Rice Quantum Institute, Rice
University, Houston, Texas 77251, USA \\
$^3$Department of Physics, Tokyo Institute of Technology,
Tokyo 152-8551, Japan \\
$^4$ERATO Macroscopic Quantum Control Project, JST, Tokyo 113-8656, Japan
}

\date{\today}

\begin{abstract}
A self-trapped Bose-Einstein condensate is shown to be stabilized in
two-dimensional free space by Rabi oscillations between two hyperfine
states which make an effective interatomic interaction oscillate in time.
The stabilization mechanism is elucidated by using a two-component
Gross-Pitaevskii equation combined with a variational analysis.
The parameter regime of stability is investigated.
\end{abstract}


\maketitle

\narrowtext

\section{Introduction}

A droplet of water is a self-trapped object, in which attractive and
repulsive interactions between the water molecules are balanced.
Such a self-trapped system is hard to achieve in the usual gaseous phase,
since thermal expansion of the gas must be prevented by a strong
attractive interaction between particles, which, however, would lead to
collapse of the gas into a denser phase.
A gaseous Bose-Einstein condensate (BEC) is the most promising candidate
for creating a self-trapped gas, or a ``BEC droplet,'' because of
unprecedented controllability of the parameters of the system.

In one dimension (1D), a BEC droplet, or a matter-wave bright soliton, is
stable with an attractive interaction between atoms counterbalancing
zero-point quantum pressure.
This novel state of a gas has been experimentally achieved by the ENS
group~\cite{Khay} and the Rice group~\cite{Strecker} by using a $^7{\rm
Li}$ condensate in quasi-1D traps.
In 2D and higher dimensions, however, there is neither a stable nor
metastable state of a BEC droplet, if the interaction is short-range and
constant in time~\cite{Sulem}.

Dynamic stabilization is a possible way to obtain a BEC droplet in 2D and
3D.
It has been shown that if the interaction is made to oscillate in time, a
BEC droplet can be dynamically stabilized in 2D~\cite{Saito03,Ab03} by a
mechanism similar to that for the stabilization of an inverted
pendulum~\cite{Feynman}.
This stabilization mechanism for a BEC droplet has been studied by several
authors~\cite{Kevrekidis,Montesinos,Gaspar,Trippen,Itin,Malomed}.
We have shown that the same mechanism can also stabilize a BEC droplet in
3D in the presence of dissipation~\cite{Saito04}.
Another dynamic scheme to stabilize a BEC droplet in 3D is to use feedback
control of the interaction~\cite{Saito06}.

In respect of the stabilization method proposed in
Refs.~\cite{Saito03,Ab03}, we must oscillate the interaction between
repulsive and attractive at a frequency much higher than the
characteristic frequencies of the system.
If we use the magnetic Feshbach resonance~\cite{Inouye} to oscillate the
interaction, we should oscillate the strength of the applied magnetic
field at such a high frequency.
We propose in the present study a method to achieve this dynamic
stabilization without using any time-dependent Feshbach control of the
interaction.
Our idea is to use two hyperfine states with different scattering
lengths.
We will show that the Rabi oscillation between these hyperfine states
causes an effective oscillation of the scattering length, thereby
stabilizing a BEC droplet through the mechanism proposed in
Refs.~\cite{Saito03,Ab03}.
This method simply irradiates the system with a constant electromagnetic
wave.

This paper is organized as follows.
Section~\ref{s:for} is devoted to describing a system of a two-component
BEC driven by an external field.
Section~\ref{s:num} performs numerical integration of the two-component
Gross-Pitaevskii equation to show that the BEC droplet is indeed
dynamically stabilized.
The stability diagram for the scattering lengths is obtained.
Section~\ref{s:var} derives an effective single-component equation, and
combines it with a variational method to elucidate the stabilization
mechanism for a BEC droplet by hyperfine Rabi oscillations.
Section~\ref{s:conc} provides conclusions.

\section{Formulation of the problem}
\label{s:for}

We consider a situation in which bosonic atoms with hyperfine degrees of
freedom are irradiated by a uniform electromagnetic wave that is resonant
with the energy difference between two hyperfine states of the atoms.
The atoms are assumed to be tightly confined in the $z$ direction by
harmonic potential $m \omega_z z^2 / 2$ that is independent of the
hyperfine state.
If $\hbar \omega_z$ is much larger than the other characteristic energy
scales, the wave function in the $z$ direction is frozen in the ground
state of the harmonic potential, and then the system can be reduced to
2D.
We will consider the 2D system in the subsequent analysis.

The Hamiltonian for the system consists of three parts:
\begin{equation} \label{H}
\hat H = \hat H_0 + \hat H_{\rm f} + \hat H_{\rm int}.
\end{equation}
The non-interacting part of the Hamiltonian $\hat H_0$ is given by
\begin{equation} \label{H0}
\hat H_0 = \int d\bm{r} \sum_{i = 1}^2 \hat\psi_i^\dagger(\bm{r})
\left[ -\frac{\hbar^2}{2m} \nabla^2 + V_i(\bm{r}) \right]
\hat\psi_i(\bm{r}),
\end{equation}
where $\hat\psi_i(\bm{r})$ annihilates an atom in hyperfine state $i$, $m$
is the atomic mass, and $V_i(\bm{r})$ is an external trapping potential
for hyperfine state $i$.
The effect of the resonant field on atomic state is described by
\begin{equation} \label{Hmw}
\hat H_{\rm f} = \hbar \int d\bm{r} \left[ \Omega
	\hat\psi_1^\dagger(\bm{r}) \hat\psi_2(\bm{r}) + \Omega^*
\hat\psi_2^\dagger(\bm{r}) \hat\psi_1(\bm{r}) \right],
\end{equation}
where $|\Omega|$ is the Rabi frequency and ${\rm arg}(\Omega)$ is the
phase of the field.
The interaction between ultracold atoms is short-range and described by
the contact Hamiltonian,
\begin{equation} \label{Hint}
\hat H_{\rm int} = \frac{2 \pi \hbar^2}{m} \sqrt{\frac{m \omega_z}{2 \pi
\hbar}} \int d\bm{r} \sum_{i, j =1}^2 a_{ij} \hat\psi_i^\dagger(\bm{r})
\hat\psi_j^\dagger(\bm{r}) \hat\psi_j(\bm{r}) \hat\psi_i(\bm{r}),
\end{equation}
where $a_{ij}$ is the $s$-wave scattering length between atoms in
hyperfine states $i$ and $j$.
The factor $[m \omega_z / (2 \pi \hbar)]^{1/2}$ in Eq.~(\ref{Hint}) comes
from the integration with respect to $z$.

In the mean-field approximation, the system at zero temperature is
described by the 2D Gross-Pitaevskii (GP) equation,
\begin{subequations} \label{GP}
\begin{eqnarray}
\label{GP1}
i \hbar \frac{\partial \psi_1}{\partial t} & = & \left(
-\frac{\hbar^2}{2m} \nabla^2 + V_1 \right) \psi_1 + \hbar \Omega \psi_2
+ g_{11} |\psi_1|^2 \psi_1 \nonumber \\
& & + g_{12} |\psi_2|^2 \psi_1,
\\
i \hbar \frac{\partial \psi_2}{\partial t} & = & \left(
-\frac{\hbar^2}{2m} \nabla^2 + V_2 \right) \psi_2 + \hbar \Omega^* \psi_1
+ g_{22} |\psi_2|^2 \psi_2 \nonumber \\
& & + g_{12} |\psi_1|^2 \psi_2,
\end{eqnarray}
\end{subequations}
where $\psi_i(\bm{r}, t)$ is the macroscopic wave function satisfying the
normalization condition,
\begin{equation}
\int d\bm{r} \sum_{i=1}^2 |\psi_i(\bm{r}, t)|^2 = N,
\end{equation}
with $N$ being the number of atoms, and
\begin{equation}
g_{ij} = \frac{4 \pi \hbar^2 a_{ij}}{m} \sqrt{\frac{m \omega_z}{2 \pi
\hbar}}
\end{equation}
representing the interaction coefficients.

\section{Stabilization of a two-component BEC droplet}
\label{s:num}

The interaction parameter of the system oscillates between $g_{11}$ and
$g_{22}$ by the Rabi oscillation between two hyperfine states.
If one of them is positive and the other is negative, the stabilization
mechanism in Refs.~\cite{Saito03,Ab03} is expected to be applicable to the
present system.
This is the main idea of the present paper.

We assume that all the atoms are initially in hyperfine state 1, and
confined in radial trapping potential $V_1 = m \omega_\perp^2 r^2 / 2$.
The initial wave function $\psi_1$ is then the ground state of GP equation
(\ref{GP1}) with $\psi_2 = 0$ and $\Omega = 0$.
At $t = 0$, the radial trapping potential is switched off and the
radiation field is switched on, i.e., $V_1 = V_2 = 0$ and $\Omega \neq 0$
for $t > 0$.

\begin{figure}[t]
\includegraphics[width=8.5cm]{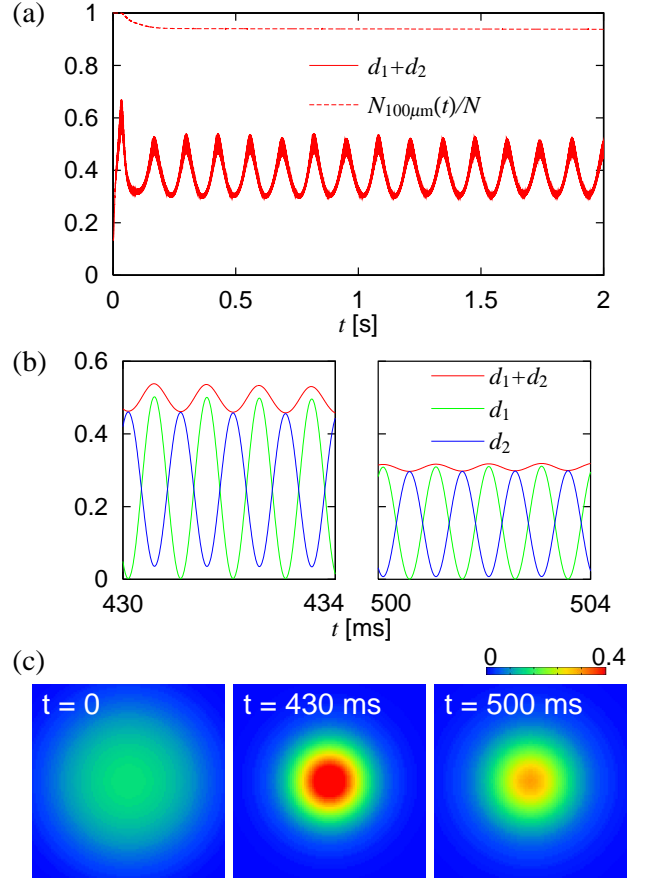}
\caption{
(Color) (a) Time evolution of the normalized peak density $d_1 +
d_2$ and that of the fraction of atoms within $r = 100$ $\mu{\rm m}$
for $a_{11} = 0.75$ nm, $a_{22} = -1.5$ nm, $a_{12} = 0$, and $\Omega = 2
\pi \times 500$ Hz.
The initial state is the ground state in the presence of a radial trapping
potential with $\omega_\perp = 2\pi \times 25$ Hz, and the trapping
potential is switched off at $t = 0$.
(b) Time scale of (a) is magnified.
Small-amplitude oscillations in (b) are invisible in (a) because of the
time resolution.
(c) Snapshots of the normalized density profiles $(|\psi_1|^2 +
|\psi_2|^2) \hbar / (m \omega_\perp N)$ at $t = 0$, 430 ms, and 500 ms.
The size of the images is 30 $\mu{\rm m}$ $\times$ 30 $\mu{\rm m}$.
}
\label{f:ev}
\end{figure}
We first investigate dynamics of the system for specific values of
scattering lengths: $a_{11} = 0.75$ nm, $a_{22} = -1.5$ nm, and $a_{12} =
0$.
The radial trapping frequency for preparing the initial state is
$\omega_\perp = 2\pi \times 25$ Hz, and the trapping frequency for
confinement in the $z$ direction is $\omega_z = 2\pi \times 5$ kHz.
The ratio $\omega_z / \omega_\perp = 200$ is sufficient for the 2D
approximation to be valid.
The number of atoms is $N = 2500$, and the Rabi frequency is
$\Omega = 2 \pi \times 500$ Hz.
The ratio $\omega_z / \Omega = 10$ assures that no axial modes are excited
by the Rabi oscillation.
For this condition, the attractive interaction by $a_{22}$ dominates the
kinetic pressure, and the system collapses even in the absence of the
trapping potential if the atomic state is fixed to the hyperfine state 2.

Figure~\ref{f:ev} (a) shows the time evolution of normalized peak density
$d_1 + d_2$ and that of the fraction of atoms around the center,
$N_{100 \mu {\rm m}}(t) / N$, where
\begin{eqnarray}
d_i(t) & = & \frac{\hbar}{m \omega_\perp N} |\psi_i(r = 0, t)|^2
\;\;\;\;\; (i = 1, 2), \\
N_\rho(t) & = & \sum_{i=1}^2 \int_0^\rho 2 \pi r |\psi_i(r, t)|^2 dr.
\end{eqnarray}
We can clearly see that collapse is prevented and the BEC droplet is
dynamically stabilized.
The Rabi oscillation between the two hyperfine states induces oscillations
of the total density at the same frequency, with the amplitude being
larger for higher density (compare the left and right panels in
Fig.~\ref{f:ev} (b)).
The long-period oscillation at the frequency of about 8 Hz persists for a
long time (solid curve in Fig.~\ref{f:ev} (a)).
The initial decrease in the dashed curve in Fig.~\ref{f:ev} (a) shows that
a small fraction of atoms are lost from the central region mainly during
the initial formation of the droplet.
These atoms have escaped from the droplet for lack of the radial trapping
potential.

\begin{figure}[t]
\includegraphics[width=9cm]{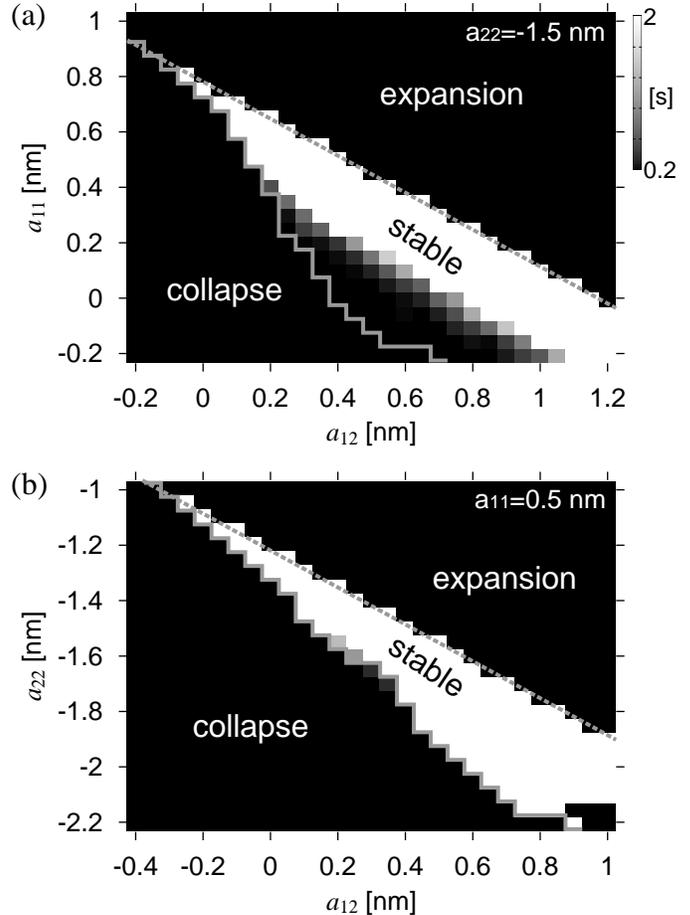}
\caption{
Stability diagrams for a BEC droplet with respect to (a) $a_{12}$ and
$a_{11}$ for $a_{22} = -1.5$ nm, and (b) $a_{12}$ and $a_{22}$ for
$a_{11} = 0.5$ nm.
The gray scale shows the lifetime of the BEC droplet.
The white region has a lifetime of at least 2 s.
The solid lines divide the regions of collapse and expansion.
The dashed lines are given by Eq.~(\ref{aline}).
}
\label{f:diagram}
\end{figure}
We next consider the stability of the BEC droplet for various values of
scattering lengths.
Figure~\ref{f:diagram} (a) shows stability diagram with respect to
$a_{11}$ and $a_{12}$ for $a_{11} = -1.5$ nm, and Fig.~\ref{f:diagram} (b)
with respect to $a_{22}$ and $a_{12}$ for $a_{11} = 0.5$ nm.
In the white region, the lifetime of the BEC droplet is at least 2
seconds.
We find that a band of stable region lies between regions of collapse and
expansion.
These two types of instability are divided by the solid lines in
Fig.~\ref{f:diagram}.
The stable region disappears for $a_{12} \lesssim -0.2$ nm.
As we can see in Fig.~\ref{f:diagram} (a), the value of $a_{11}$ in the
stable region become negative as $a_{12}$ increases.
However, this would not hinder an initial state of the condensate from
being prepared.
One can prepare the condensate in a positive $a_{11}$ and switch $a_{11}$
to an appropriate negative value at $t = 0$.

The stability region is not sensitive to the Rabi frequency $\Omega$.
To see this, we normalize the time, length, and wave function by
$|\Omega|^{-1}$, $[\hbar / (m |\Omega|)]^{1/2}$, $N [m |\Omega| /
\hbar]^{1/2}$, respectively.
GP equation (\ref{GP1}) then becomes
\begin{eqnarray} \label{GPn}
i \frac{\partial \psi_1}{\partial t} & = & \left( -\frac{\nabla^2}{2} +
\frac{\omega_\perp^2}{2 |\Omega|^2} r^2 \right) \psi_1 + \psi_2
\nonumber \\
& & + 4 \pi N \sqrt{\frac{m \omega_z}{2 \pi \hbar}} \left( a_{11} |\psi_1|^2
\psi_1 + a_{12} |\psi_2|^2 \psi_1 \right).
\nonumber \\
\end{eqnarray}
We note that $\Omega$ is included only in the term of the radial trapping
potential used for preparing the initial state, and therefore the change
in $\Omega$ only affects the initial wave function in the normalized form
of GP equation~(\ref{GPn}).
We performed numerical simulations also for $\Omega = 2 \pi \times 1$ kHz,
and confirmed that the stability region is almost the same as in
Fig.~\ref{f:diagram}.

\section{Analytic results}
\label{s:var}

In this section, we analytically study the stabilization of a BEC droplet
shown in Sec.~\ref{s:num}.

We first transform wave functions $\psi_1$ and $\psi_2$ into
\begin{subequations} \label{Psi}
\begin{eqnarray}
\Psi_1 & = & \psi_1 \cos |\Omega| t + i e^{i \phi} \psi_2 \sin |\Omega| t,
\\
\Psi_2 & = & \psi_2 \cos |\Omega| t + i e^{-i \phi} \psi_1 \sin |\Omega|
t,
\end{eqnarray}
\end{subequations}
where $\Omega = e^{i\phi} |\Omega|$.
Substituting Eq.~(\ref{Psi}) into GP equation (\ref{GP}), we find that
the terms proportional to $|\Omega|$ vanish, and we obtain
\begin{widetext}
\begin{eqnarray}
i \hbar \frac{\partial \Psi_1}{\partial t} & = & -\frac{\hbar^2}{2m}
\nabla^2 \Psi_1 + \frac{g_{11} + g_{22}}{8} \left( 3 |\Psi_1|^2 \Psi_1 + 2
|\Psi_2|^2 \Psi_1 - e^{2i\phi} \Psi_1^* \Psi_2^2 \right)
\nonumber \\
& & + \frac{g_{12}}{4} \left( |\Psi_1|^2 \Psi_1 + 2 |\Psi_2|^2 \Psi_1 +
e^{2i\phi} \Psi_1^* \Psi_2^2 \right)
\nonumber \\
& & + (g_{11} - g_{22}) \left\{ \frac{\cos 2|\Omega| t}{2} |\Psi_1|^2
\Psi_1 + \frac{i \sin 2|\Omega| t}{4} \left[ e^{-i\phi} \Psi_2^* \Psi_1^2
- e^{i\phi} \left( 2 |\Psi_1|^2 + |\Psi_2|^2 \right) \Psi_2 \right]
\right\}
\nonumber \\
& & + \frac{g_{11} + g_{22} - 2 g_{12}}{8} \bigl\{ \cos 4|\Omega| t \left(
|\Psi_1|^2 \Psi_1 - 2 |\Psi_2|^2 \Psi_1 + e^{2i\phi} \Psi_1^* \Psi_2^2
\right)
\nonumber \\
& & + i \sin 4|\Omega| t \left[ e^{-i\phi} \Psi_2^* \Psi_1^2 + e^{i\phi}
\left( -2 |\Psi_1|^2 + |\Psi_2|^2 \right) \Psi_2 \right] \bigr\},
\label{dPsi1dt}
\end{eqnarray}
\end{widetext}
where we set $V_1 = V_2 = 0$.
The expression of $i \hbar \partial \Psi_2 / \partial t$ is obtained by
replacements $\Psi_1 \leftrightarrow \Psi_2$, $g_{11} \leftrightarrow
g_{22}$, and $e^{i\phi} \leftrightarrow e^{-i\phi}$ in
Eq.~(\ref{dPsi1dt}).

The initial state of $\psi_2$ is assumed to be $\psi_2 = 0$ and then
$\Psi_2 = 0$ at $t = 0$.
Setting $\Psi_2 = 0$ in the expression of $i \hbar \partial \Psi_2 /
\partial t$ gives
\begin{eqnarray}
i \hbar \frac{\partial \Psi_2}{\partial t} & = & \frac{i e^{-i\phi}}{8}
\bigl[ 2 (g_{11} - g_{22}) \sin 2|\Omega| t \nonumber \\
& & + (g_{11} + g_{22} - 2 g_{12}) \sin
4|\Omega| t \bigr] |\Psi_1|^2 \Psi_1,
\end{eqnarray}
which indicates that $\Psi_2$ is of the order of $1 / |\Omega|$.
We therefore approximate that $\Psi_2$ is always zero.
Equation~(\ref{dPsi1dt}) is then approximately reduced to the
single-component GP equation,
\begin{equation} \label{Psi1GP}
i \hbar \frac{\partial \Psi_1}{\partial t} = -\frac{\hbar^2}{2m} \nabla^2
\Psi_1 + G(t) |\Psi_1|^2 \Psi_1,
\end{equation}
where
\begin{equation}
G(t) = G_0 + G_1 \cos 2|\Omega| t + G_2 \cos 4|\Omega| t,
\end{equation}
with
\begin{eqnarray}
\label{G0}
G_0 & = & \frac{3 (g_{11} + g_{22}) + 2 g_{12}}{8}, \\
G_1 & = & \frac{g_{11} - g_{22}}{2}, \\
G_2 & = & \frac{g_{11} + g_{22} - 2 g_{12}}{8}.
\end{eqnarray}

To check if Eq.~(\ref{Psi1GP}) captures the stabilization mechanism for a
BEC droplet, we perform a variational analysis by using a Gaussian
variational wave function~\cite{Garcia},
\begin{equation} \label{gauss}
\Psi_1 = \frac{\sqrt{N}}{\sqrt{\pi} R(t)} \exp\left[ -\frac{r^2}{2 R^2(t)}
	+ i \frac{m \dot{R}(t) r^2}{2 \hbar R(t)} \right],
\end{equation}
where $R(t)$ is the variational parameter that characterizes the size of
the droplet, and the second term in the exponent describes mass current
which is required to satisfy the equation of continuity.
The action that derives Eq.~(\ref{Psi1GP}) has the form,
\begin{equation} \label{action}
K = \int d \bm{r} dt \left[ -i \hbar \Psi_1^* \frac{\partial}{\partial t}
\Psi_1 - \frac{\hbar^2}{2m} \Psi_1^* \nabla^2 \Psi_1 + \frac{G(t)}{2}
|\Psi_1|^4 \right].
\end{equation}
Substituting Eq.~(\ref{gauss}) into Eq.~(\ref{action}) and taking $\delta
K / \delta R = 0$, we obtain the equation of motion for $R(t)$,
\begin{equation} \label{eom}
\ddot{R}(t) = \left[ \frac{\hbar^2}{m^2} + \frac{N G(t)}{2\pi m} \right]
\frac{1}{R^3(t)}.
\end{equation}
Noting that the dynamics comprises rapid and slow oscillations as shown in
Figs.~\ref{f:ev} (a) and \ref{f:ev} (b), we separate $R(t)$ into a slowly
varying part $R_0(t)$ and rapidly oscillating part $\varepsilon(t)$ as
\begin{equation}
R(t) = R_0(t) + \varepsilon(t).
\end{equation}
The rapidly oscillating part of Eq.~(\ref{eom}) is approximated to be
\begin{equation} \label{epsdot}
\ddot{\varepsilon}(t) \simeq \frac{N}{2\pi m R_0^3(t)} \left( G_1 \cos
2|\Omega| t + G_2 \cos 4|\Omega| t \right),
\end{equation}
where we neglect the terms of order $|\Omega|^{-2}$.
Since $R_0$ can be regarded as a constant in the fast time scale in
Eq.~(\ref{epsdot}), we have
\begin{equation} \label{eps}
\varepsilon(t) \simeq -\frac{N}{32\pi m R_0^3(t) |\Omega|^2} \left( 4 G_1
\cos 2|\Omega| t + G_2 \cos 4|\Omega| t \right).
\end{equation}
Substituting Eq.~(\ref{eps}) into the slowly varying part of
Eq.~(\ref{eom}),
\begin{eqnarray}
\ddot{R}_0(t) & = & \left( \frac{\hbar^2}{m^2} + \frac{N G_0}{2 \pi m}
\right) \frac{1}{R_0^3(t)} \nonumber \\
& & - \frac{3 N}{2 \pi m R_0^4(t)} \overline{\varepsilon(t) \left( G_1
\cos 2|\Omega| t + G_2 \cos 4|\Omega| t \right)} \nonumber \\
& & + O(\varepsilon^2),
\end{eqnarray}
where $\overline{\cdots}$ indicates time average for the rapid oscillation,
we finally obtain an effective equation of motion for slowly varying
parameter $R_0(t)$ as
\begin{equation} \label{eff}
\ddot{R}_0(t) \simeq \left( \frac{\hbar^2}{m^2} + \frac{N G_0}{2 \pi m}
\right) \frac{1}{R_0^3(t)} + \frac{3 N^2 \left( 4 G_1^2 + G_2^2
\right)}{128 \pi m^2 |\Omega|^2 R_0^7(t)}.
\end{equation}

The first term on the right-hand side of Eq.~(\ref{eff}) originates from
the kinetic energy and the constant part of the interaction energy.
For stationary $R_0$ to exist, the coefficient of the first term must be
negative, i.e.,
\begin{equation} \label{ineq}
\frac{m N G_0}{2 \pi \hbar^2} < -1,
\end{equation}
which gives the condition for stability against expansion.
Using Eqs.~(\ref{G0}) and (\ref{ineq}), the boundary between the stable
and unstable regions is given by
\begin{equation} \label{aline}
3 (a_{11} + a_{22}) + 2 a_{12} = -\frac{4}{N} \sqrt{\frac{2 \pi \hbar}{m
\omega_z}}.
\end{equation}
This boundary is shown in Fig.~\ref{f:diagram} by the dashed lines, which
show close agreement with the numerically obtained boundary.

We note that the second term on the right-hand side of Eq.~(\ref{eff}),
which originates from the oscillating part of the interaction in
Eq.~(\ref{Psi1GP}), prevents the system from collapsing.
Thus, the effective oscillation of the interaction by the Rabi
oscillation dynamically stabilizes the BEC droplet.
The system is stationary ($\ddot{R}_0 = 0$) with the size of the droplet,
\begin{equation} \label{Rs}
R_0 = \left[ \frac{3 N^2 (4 G_1^2 + G_2^2)}{64 \pi |\Omega|^2 \left| 2 \pi
\hbar^2 + m N G_0 \right|} \right]^{1/4},
\end{equation}
which is $\simeq 6.4$ $\mu{\rm m}$ for the parameters in Fig.~\ref{f:ev}.
Comparing the density profile at $t = 0$ [$R = (\hbar / m
\omega_\perp)^{1/2} \simeq 7.6$ $\mu{\rm m}$] with those at $t = 430$ and
500 ms in Fig.~\ref{f:ev} (c), we can see that this value of $R_0$ agrees
well with the numerical one.
The frequency of the small oscillation around Eq.~(\ref{Rs}) is obtained
as
\begin{equation}
\omega = \frac{8 \sqrt{2} |\Omega| \left| 2 \pi \hbar^2 + m N G_0
\right|}{\sqrt{3} m N \sqrt{4 G_1^2 + G_2^2}},
\end{equation}
which is $\sim 2\pi \times 12$ Hz for the parameters in Fig.~\ref{f:ev},
in qualitative agreement with the numerical result of about $2\pi \times
8$ Hz.
The difference may be due to the large amplitude of the oscillation as
shown in Fig~\ref{f:ev} (a).

\section{Conclusions and discussion}
\label{s:conc}

We have studied a 2D BEC of atoms in two hyperfine states, which undergo
Rabi oscillation by an external field.
When scattering lengths $a_{11}$, $a_{22}$, and $a_{12}$ are different,
the Rabi oscillation leads to effective oscillation of the interaction.
If the scattering lengths fall in the stable regions such as those in
Fig.~\ref{f:diagram}, the self-trapped BEC in free space --- a BEC droplet
--- can be stabilized by the mechanism discussed in
Ref.~\cite{Saito03,Ab03}.

Solving the two-component GP equation (\ref{GP}) numerically, we have
shown that the stabilization mechanism works and the BEC droplet is
stabilized (Fig.~\ref{f:ev}).
The stabilization is shown to be possible even if the system starts from
the ground state in a radial trapping potential, which is switched off at
$t = 0$.
We have obtained the stability diagram with respect to scattering lengths
(Fig.~\ref{f:diagram}), in which there exists the stability region between
the regions of collapse and expansion.

To elucidate the mechanism for the dynamic stabilization of a BEC droplet,
we approximately reduced the two-component GP equation to a single
component equation (\ref{Psi1GP}) and studied it by using the Gaussian
variational method.
We found that the Rabi oscillation generates an effective potential that
counteracts the attractive interaction, thereby creating a stable droplet
state.
The size of the droplet and its breathing-mode oscillation frequency are
obtained by the variational analysis, which produces results in
qualitative agreement with the numerical ones.

Our scheme applies to any system where atoms in the two relevant hyperfine
states are not transferred to other states by inelastic collisions and the
coupling field is available that makes Rabi oscillations between the two
states.
For example, the hyperfine states $|F, m_F \rangle = |1, 1 \rangle$ and
$|2, 2 \rangle$ of $^{87}{\rm Rb}$ and $^{23}{\rm Na}$, and $|2, 2
\rangle$ and $|3, 3 \rangle$ of $^{85}{\rm Rb}$ are possible candidates,
if their scattering lengths can be tuned to fall in the stable region.
Unfortunately, for $|1, 1 \rangle$ and $|2, 2 \rangle$ hyperfine states of
$^7{\rm Li}$, the scattering lengths are $a_{11} \simeq 0.26$
nm~\cite{Strecker}, $a_{22} \simeq -1.4$ nm~\cite{Abraham}, and $a_{12}
\simeq -1.4$ nm~\cite{McAlexander}, and this set of scattering lengths is
located out of the stability region.
By systematic calculations as in Fig.~\ref{f:diagram}, we found that there
is no stable set of $a_{11}$ and $a_{22}$ for this value of $a_{12}$.

Using the above pairs of hyperfine states, inelastic decay of the upper
hyperfine state occurs only through dipole-dipole interaction.
The dipolar decay rate $K$ is typically less than $10^{-14}$ ${\rm cm}^3 /
{\rm s}$~\cite{Boesten} and the peak density in Fig.~\ref{f:ev} (a) is
$n \simeq 0.4 N m \omega_\perp / \hbar [m \omega_z / (\pi \hbar)]^{1/2}$,
which gives $n K \lesssim 0.2$ ${\rm s}^{-1}$.
Therefore, the lifetime of the droplet in the presence of the dipolar loss 
is estimated to be more than 5 s.

Another possibility is the use of two states in the same hyperfine
manifold, e.g., $|1, 1 \rangle$ and $|1, 0 \rangle$, in which modest
magnetic fields ($\sim$ 1-10 G) should be applied to suppress the
spin-exchange collisions ($|1, 0 \rangle + |1, 0 \rangle \rightarrow |1, 1
\rangle + |1, -1 \rangle$).
For example, for these states of $^7{\rm Li}$, there are regions in which
Eq.~(\ref{aline}) becomes small and negative as a function of magnetic
field, suggesting that the stabilization is possible.

The 2D potential may not be perfectly flat in experiments, and a residual
weak potential in 2D, e.g., a weak harmonic potential, may exist.
Even in this case, we can distinguish self trapping from trapping by a
residual potential, if the size of a droplet is made much smaller than the
scale of inhomogeneity in the 2D potential.

As in the case of oscillating the scattering length~\cite{Saito03,Ab03},
we have not been able to create a stable BEC droplet in 3D in the absence
of dissipation.
One reason for difficulty is that collapse is stronger for higher
dimensions~\cite{Sulem}, and therefore the amplitude of the rapid
oscillation should be larger to suppress the collapse.
However, this breaks the small-amplitude and Gaussian approximations, and
unstable modes are dynamically induced.

The present method is different from the one in Ref.~\cite{Saito03,Ab03},
in that explicit oscillation of the interaction using the Feshbach
resonance is not needed.
The rapid modulation of the applied magnetic field for the Feshbach
resonance may cause some experimental difficulties, e.g., an induced
current in the apparatus and resultant heating, and therefore the present
method is expected to facilitate the experimental realization of a BEC
droplet.

\begin{acknowledgments}
This work was supported by Grants-in-Aid for Scientific Research (Grant
Nos.\ 17740263 and 17071005) and by the 21st Century COE programs on
``Coherent Optical Science'' and ``Nanometer-Scale Quantum Physics'' from
the Ministry of Education, Culture, Sports, Science and Technology of
Japan.
The work at Rice was supported by the NSF, ONR, and the Welch Foundation
(C-1133).
\end{acknowledgments}

\end{document}